\documentclass [12pt]{article}
\usepackage {graphicx}
\usepackage {longtable}

\begin{document}

\bigskip

\section*{A Conformally Invariant Approach to Estimation  
of Relations Between Physical Quantities}

\bigskip

\subsubsection*{M.V. Gorbatenko and G.G. Kochemasov}

\bigskip

\begin{center}
Russian Federal Nuclear Center - All-Russian Research Institute of 
Experimental Physics, Sarov, Nizhni Novgorod region, Russia 
\end{center}

\bigskip

\begin{center}
\textbf{Abstract}
\end{center}

A.V.Pushkin's approach based on conformal geometrodynamics (CGD) to 
calculation of quantitative relations between physical quantities is 
presented and analyzed. In the simplest cases of the stationary solutions to 
the CGD equations the approach implies separation of internal and external 
parts (relative to a certain boundary) from the solutions and using inverse 
transformations transforming the parts into each other. For the 
quasi-stationary (metastable) states, the possibility of the nonperturbative 
calculation of their lifetimes is shown. The approach is illustrated by 
several examples. In particular, it is shown that the Dirac ``large number 
hypothesis'' is a consequence of the approach. Also, the evaluated radiation 
lifetime of the first excited level of 2p hydrogen atom and neutron lifetime 
are presented.

\bigskip

\section*{1. Introduction}

\bigskip

This paper addresses a conformally invariant approach to the estimation of 
relations between fundamental physical quantities. As far as we know, 
A.V.Pushkin was the first to employ the approach (see, e.g., [1], [2]), that 
is why we will name the approach after him. Unfortunately, the approach 
itself is presented fragmentarily, without appropriate elucidation in the 
literature. The objective of this paper is to bridge the gap.

Eventually, Puskin's approach is based on the analysis of symmetry of the 
conformal geometrodynamics equations with energy-momentum tensor of purely 
geometric nature and symmetry of the conformal quantum field theory with the 
same tensor as a vacuum polarization tensor. According to Pushkin, in the 
quantum field theory a symmetry group termed the "Monster" group acts. 
Reasoning from the above considerations, ref. [1] estimates fine structure 
constant $\alpha $ and proton to electron mass ratio ${{m_{P}}  
\mathord{\left/ {\vphantom {{m_{P}}  {m_{e}} }} \right. 
\kern-\nulldelimiterspace} {m_{e}} }$ and ref. [2] evaluates the background 
radiation temperature.

For convenience of the reader, to whom papers [1], [2] may be unavailable, 
present the net result of these works. Thus, according to Pushkin, the fine 
structure constant is

\[
\alpha _{theory}^{ - 1} = \frac{{\bar {N} + \Delta} }{{dim\Omega _{2}} } = 
137.03599079... \quad ,
\]

\noindent
where $\bar {N} = N_{tot} - N_{1} = 274$, $N_{tot} = \sum\limits_{i = 1}^{6} 
{N_{i}}  $=286, $N_{i} $ is the total number of Killing vectors for all 
$i$-th order subgroups of the 15-parameter conformal group, $\Delta = \left( 
{Ñ_{M} - 1} \right)^{ - \frac{{1}}{{2}}}$ is the quantum anomalous 
dimension, $C_{M} = \bar {N} - N_{2} = 194$ is the amount of the Monster 
conjugacy classes, $dim\Omega _{2} $=2 is merely the dimension of 2D 
surfaces.

The following relation is valid for the proton to electron mass ratio: 

 $\quad
\frac{{m_{p}} }{{m_{e}} } = \frac{{d_{B}} }{{120 - N_{1}} } + 
\frac{{\frac{{\bar {N}}}{{2}}\left[ {N_{tot} - \frac{{\bar {N}}}{{2}}} 
\right]}}{{\left( {N_{1} - 4} \right)C_{M}} }=$ 1836.1527..,

\noindent
where $_{} d_{B} $=196884 is the dimension of Griess algebra. The 
reciprocal of this dimension controls the accuracy of the calculation of 
$\Delta $ and $\left( {N_{1} - 4} \right)C_{M} $ in the first and second 
formulas, respectively, and, as a consequence, of the ultimate results. 

Ref. [2] calculates the ratio of temperature of cosmological (relict) 
radiation to electron rest energy, so that the temperature is 

 $T = k^{ - 1}\frac{{m_{e} c^{2}}}{{\rho S_{Mac}} }=$ 2.736 $K$,

\noindent
where $\rho $=696729600 is the number of Weyl group elements, viz. the 
symmetry group of lattice $E_{8} $, $S_{Mac} = 7 \cdot \frac{{1}}{{3}} \cdot 
2 \cdot \frac{{2}}{{3}} = \frac{{28}}{{9}}$ is a characteristic of the first 
excited state of the internal space characterized by the lattice $E_{8} $. 
In view of the aforesaid, no relation presented includes any 
phenomenological (adjustable) parameters; only algebraic characteristics of 
the conformal group and Monster group enter into the relations, with the 
characteristics of these groups relating with each other.

Discussions of the papers by Pushkin suggest that it is the estimations of 
the physical quantities and relations between them that provoke, on the one 
hand, the greatest interest and, on the other hand, most emotional sentences 
and questions. Therefore it seems reasonable to try to present in a form as 
much systematized as possible at least some arguments and considerations 
needed to understand Pushkin's approach to the estimation of relations 
between physical quantities. The arguments and considerations make up a 
certain ``construct''. According to Pushkin, it is the presence of this 
construct that differentiates his method from a formal numerology, which is 
a manipulation of numbers in order to obtain needed relations. Besides, it 
seems reasonable to illustrate the application of Pushkin's method with 
several specific examples. 

That intention proved hard to implement. Pushkin's method includes a wide 
range of techniques, among which by no means all can be explained by us. Our 
way to overcome this obstacle is to restrict ourselves only to the part of 
the computational technique which is clear to us. Actually, this means that 
we restrict our consideration to that analysis part, which accounts only for 
the CGD equation symmetry properties corresponding to the conformal 
transformations and differentiable changes of coordinates. In so doing the 
properties governed by the Monster group are not taken into consideration 
explicitly. Besides, according to the data available to us, after the 
untimely death of Pushkin in 2004 no systematized statement of his method 
was left, therefore at some points we will have to conjecture for the 
author. The non-author version of presentation will inevitably include 
subjective points stemming from the differences in understanding of the 
approach under discussion. So Pushkin's method version suggested in this 
paper should not be viewed as the only possible. Despite these reservations, 
the paper may prove helpful in the attempt to understand the outputs placed 
by Pushkin in his papers as well as unpublished ones which Pushkin discussed 
with his colleagues.

\section*{2. Conformally inverse transformations of the CGD equations}

\bigskip

\subsection*{2.1. Conformal geometrodynamics equations}

\bigskip

In this paper by the conformal geometrodynamics (CGD) is meant the theory 
based on equation [3]

\begin{equation}
\label{eq1}
\left. {\begin{array}{l}
 {R_{\alpha \beta}  - \frac{{1}}{{2}}g_{\alpha \beta}  = T_{\alpha \beta} }  
\\ 
 {T_{\alpha \beta}  \equiv - 2A_{\alpha}  A_{\beta}  - g_{\alpha \beta}  
A^{2} - 2g_{\alpha \beta}  A^{\varepsilon} _{;\varepsilon}  + A_{\alpha 
;\beta}  + A_{\beta ;\alpha} }  \\ 
 \end{array}}  \right\}.
\end{equation}

\noindent
where $R_{\alpha \beta}  $ is the Ricci tensor in four-dimensional 
Riemannian space; the semicolon means the covariant differentiation 
performed using Christoffel symbols. Tensor $T_{\alpha \beta}  $ that is 
typically associated with the matter energy-momentum tensor is determined in 
this case by Weyl vector field $A_{\alpha}  $.

Equations (\ref{eq1}) are self-similar, which reflects the absence of any absolutely 
dimensional scale in the broad variety of effects described by them, with 
the gauge invariance being local: equations (\ref{eq1}) are invariant under 
coordinate-dependent conformal transformations

\begin{equation}
\label{eq2}
g_{\alpha \beta}  \left( {x} \right) \to {g}'_{\alpha \beta}  \left( {x} 
\right) = g_{\alpha \beta}  \left( {x} \right) \cdot \phi \left( {x} 
\right),\;\;A_{\alpha}  \to {A}'_{\alpha}  = A_{\alpha}  - \frac{{\partial 
_{\alpha}  ln\phi \left( {x} \right)}}{{2\partial x^{\alpha} }}.
\end{equation}

Weyl himself originally interpreted vector $A_{\alpha}  $ as an 
electromagnetic field potential, which led to the well-known criticism on 
the part of Einstein. There is, however, another approach to the 
understanding of the physical meaning of gauge vector $A_{\alpha}  $, i.e. 
the one discussed in refs. [4]-[9], [10], [11] and elsewhere. In this 
approach the meaning of vector $A_{\alpha}  $ is determined through analysis 
of the solutions to equations (\ref{eq1}) with using no a priori assumptions. It is 
just this approach that we will adhere to in this paper.

\subsection*{2.2. Explanations to the conformal inverse transformations}

\bigskip

The conformal inverse transformations refer to the category of 
transformations (\ref{eq2}) and can be applied to stationary solutions of the CGD 
equations. Illustrate the action of the conformal inverse transformations by 
the example of the conformally flat solutions.

We start from the Minkowski space with zero Weyl vector. Given Cartesian 
coordinates in this space, the metric is

\begin{equation}
\label{eq3}
\eta _{\alpha \beta}  = diag\left( { - 1,1,1,1} \right).
\end{equation}

Consider the simplest type of the inverse conformal transformations - the 
ones that include a combination of the following two transformations:

(\ref{eq1}) Inverse coordinate transformation $\left\{ {x^{\alpha} } \right\} \to 
\left\{ {{x}'^{\alpha} } \right\}$:

\begin{equation}
\label{eq4}
x^{\alpha}  \to {x}'^{\alpha}  = \frac{{a^{2}}}{{r^{2}}}x^{\alpha} ,
\end{equation}

\noindent
where $a$ is a parameter having dimension of length, $r^{2} \equiv \eta 
_{\alpha \beta}  x^{\alpha} x^{\beta} $. From (\ref{eq4}) it follows that

\begin{equation}
\label{eq5}
x^{\alpha}  = \frac{{a^{2}}}{{{r}'^{2}}}{x}'^{\alpha} ,
\end{equation}

\noindent
where ${r}'^{2} \equiv \eta _{\alpha \beta}  {x}'^{\alpha} {x}'^{\beta}  = 
{{a^{4}} \mathord{\left/ {\vphantom {{a^{4}} {r^{2}}}} \right. 
\kern-\nulldelimiterspace} {r^{2}}}$. If originally the Riemann tensor and 
vector $A_{\alpha}  $ were zero, then upon transformations (\ref{eq4}), (\ref{eq5}) the 
Riemann tensor and vector $A_{\alpha}  $ remain zero as well. In these 
transformations the metric tensor, of course, changes. The new metric tensor 
${g}'^{\alpha \beta} $ depends on coordinates ${x}'^{\alpha} $ and is 
determined with formula

\begin{equation}
\label{eq6}
{g}'^{\alpha \beta}  = \frac{{\partial {x}'^{\alpha} }}{{\partial x^{\mu 
}}}\frac{{\partial {x}'^{\beta} }}{{\partial x^{\nu} }}\eta ^{\mu \nu} .
\end{equation}

From (\ref{eq4}) it follows that

\begin{equation}
\label{eq7}
\frac{{\partial {x}'^{\alpha} }}{{\partial x^{\beta} }} = 
\frac{{a^{2}}}{{r^{2}}}\delta _{\beta} ^{\alpha}  - 
2\frac{{a^{2}}}{{r^{4}}}x^{\alpha} \eta _{\beta \nu}  x^{\nu} .
\end{equation}

Substitution of (\ref{eq7}) into (\ref{eq6}) results in

\begin{equation}
\label{eq8}
{g}'^{\alpha \beta}  = \frac{{a^{4}}}{{r^{4}}}\eta ^{\alpha \beta} ,\quad 
{g}'_{\alpha \beta}  = \frac{{r^{4}}}{{a^{4}}}\eta _{\alpha \beta}  .
\end{equation}

In terms of primed coordinates we have:

\begin{equation}
\label{eq9}
{g}'_{\alpha \beta}  = \frac{{a^{4}}}{{{r}'^{4}}}\eta _{\alpha \beta}  .
\end{equation}

(\ref{eq2}) The dilaton conformal transformation of the following form:

\begin{equation}
\label{eq10}
{g}'_{\alpha \beta}  \to \hat {{g}'}_{\alpha \beta}  = {g}'_{\alpha \beta}  
\cdot \frac{{{r}'^{4}}}{{a^{4}}} = \eta _{\alpha \beta}  .
\end{equation}

The conformal factor in the transition from ${g}'_{\alpha \beta}  $ to $\hat 
{{g}'}_{\alpha \beta}  $ is

\begin{equation}
\label{eq11}
\phi \left( {{x}'} \right) = \left( {\sigma \left( {{x}'} \right)} 
\right)^{2} = \frac{{{r}'^{4}}}{{a^{4}}}.
\end{equation}

As a result of transformations (\ref{eq4}), (\ref{eq9}) we arrive at a space with metric 
(\ref{eq3}). The point with coordinates $\left\{ {x^{\alpha} } \right\}$ has 
transferred to the point with coordinates $\left\{ {{x}'^{\alpha} } 
\right\}$ which is calculated by formula (\ref{eq4}). 

Despite the fact that upon the transformations the form of the space metric 
remained unchanged, nevertheless, one significant change occurred: a nonzero 
field of vector $A_{\alpha}  $ appeared. The field of vector $A_{\alpha}  $ 
is given by

\begin{equation}
\label{eq12}
A_{\alpha}  \left( {{x}'} \right) = - \frac{{\left( {\sigma \left( {{x}'} 
\right)} \right)_{,\alpha} } }{{\sigma \left( {{x}'} \right)}} = - 2\eta 
_{\alpha \mu}  \frac{{{x}'^{\mu} }}{{{r}'^{2}}}.
\end{equation}

It can be shown that on the substitution of the expression for vector 
$A_{\alpha}  $ in form (\ref{eq12}) the energy-momentum tensor for the CGD equation 
vanishes. Thus, as a result of the inverse conformal transformation of the 
simplest type we arrive at a plane space, in which vector $A_{\alpha}  $ 
appears as a gradient and the energy-momentum tensor remains zero. 

Emphasize once again that the conformally plane spaces are exact solutions 
to the CGD equations that involve no approximations.

The inverse conformal transformations are not exhausted with the ones of the 
form considered in this item. A wider class of the transformations is 
discussed in ref. [2]. 

Note on terminology. In many papers by the conformally plane space is meant 
the space produced from Minkowski space merely using the conformal 
transformation. Upon this transformation the metric differs from that of the 
Minkowski space, therefore thus produced space is not a conformally plane 
space in our treatment. 

Also note that the procedure of introduction of the nonzero Weyl field and 
inverse conformal transformations admits a generalization consisting in 
replacement of the Minkowski space with an arbitrary Riemannian 
(pseudo-Riemannian) space.

\section*{3. Stationary solutions and Dirac ``large number hypothesis''}

\bigskip

Consider the issue of properties of the transformation described by the 
static spherically symmetric solution of CGD equations. We will use the term 
of particle with meaning by it a region localized in space, inside and 
outside of which fields $g_{\alpha \beta}  \left( {x} \right)$, $A_{\alpha}  
\left( {x} \right)$ are described by the branches of the general static 
spherically symmetric solution to the CGD equations. We proceed from the 
assumption that the solution is regular over the entire space, excluding the 
appearing discontinuity surface. This solution type can exist because, 
first, there are no connections to initial data in setting up the Cauchy 
problem for the CGD equations and, second, the velocity of perturbation 
propagation in continuum described by geometrodynamic energy-momentum tensor 
(\ref{eq1}) can be as fast as light velocity. For a more detailed discussion of this 
issue see ref. [13]. 

According to ref. [13], there are three types of the general solution to the 
static spherically symmetric problem, each of which is described by five 
integration constants. Some of the constants can be assumed zero. In the 
simplest case the particle is characterized only by two constants: 
gravitational radius $\left( {{{Gm} \mathord{\left/ {\vphantom {{Gm} 
{c^{2}}}} \right. \kern-\nulldelimiterspace} {c^{2}}}} \right)$ and radius 
of the Universe. Whatever the constant set, however, we obtain the solution 
having singularity for some value of radial variable $z$.

In the range of small values of the radial variable the discontinuous 
solution is close to the de Sitter solution known in the general relativity, 
while in the range of large ones to the Schwarzschild solution in the space 
coinciding asymptotically with the de Sitter space. At a certain value of 
the radial variable these two solution branches get sewn. Denote the sewing 
surface radius by $a$. A specific value of the $a$ is determined from 
additional physical considerations (see [13]), which we will not analyze 
here. To us only the fact itself of the discontinuity surface existence will 
be important. 

Constants $\left( {{{Gm} \mathord{\left/ {\vphantom {{Gm} {c^{2}}}} \right. 
\kern-\nulldelimiterspace} {c^{2}}}} \right)$, $\left( {{{c} \mathord{\left/ 
{\vphantom {{c} {H}}} \right. \kern-\nulldelimiterspace} {H}}} \right)$ can 
be interpreted as follows. For the value of the radial variable $z$ equal to 
$\left( {{{Gm} \mathord{\left/ {\vphantom {{Gm} {c^{2}}}} \right. 
\kern-\nulldelimiterspace} {c^{2}}}} \right)$, the external part of the 
solution would become singular, if it were continued to the range of small 
$z$. For the value of the radial variable $z$ equal to $\left( {{{c} 
\mathord{\left/ {\vphantom {{c} {H}}} \right. \kern-\nulldelimiterspace} 
{H}}} \right)$, the internal part of the solution would become singular, if 
it were continued to the range of large $z$.

If the described situation takes place, then a conformal transformation 
exists that swaps the internal and external solution parts with the 
discontinuity surface remaining unchanged. This transformation refers to the 
category of inverse invariant transformations. Length $a$ should therewith 
satisfy the ``golden'' section rule

\begin{equation}
\label{eq13}
a = Const \cdot \sqrt {\left( {{{Gm} \mathord{\left/ {\vphantom {{Gm} 
{c^{2}}}} \right. \kern-\nulldelimiterspace} {c^{2}}}} \right) \cdot \left( 
{{{c} \mathord{\left/ {\vphantom {{c} {H}}} \right. 
\kern-\nulldelimiterspace} {H}}} \right)} = Const \cdot \sqrt 
{\frac{{Gm}}{{cH}}} .
\end{equation}

Constant $Const$ that has appeared in the expression for $a$ is a number 
close to unity. Here we will not take up the calculation of the $Const$. We 
only note that its value is related to the amount of the discontinuity 
surface implementation methods and, evidently, differs for different 
particles. It will be assumed equal to unity except as otherwise noted. 

The obtained value of length $a$ is the discontinuity surface radius for the 
simplest spherical symmetric static particle. However, no particles offering 
the above properties exist in the Nature. All nonzero mass particles possess 
spin, electric charge and other quantum numbers and, of course, cannot be 
described by the considered spherically symmetric static solution. 

\subsection*{Example 1}

In the zeroth approximation we can neglect the presence of spin in electron 
and identify the sewing surface radius $a$ with the classic electron radius 
$\left( {{{e^{2}} \mathord{\left/ {\vphantom {{e^{2}} {mc^{2}}}} \right. 
\kern-\nulldelimiterspace} {mc^{2}}}} \right)$, i.e. assume that

\begin{equation}
\label{eq14}
\sqrt {\frac{{Gm}}{{cH}}} \approx \frac{{e^{2}}}{{mc^{2}}}.
\end{equation}

It turns out that equality (\ref{eq14}) obtained from the identification is neither 
more nor less than the relation of the Dirac ``large number hypothesis''. 
Indeed, the relation is written, as a rule, in the form of equality of two 
large numbers $N_{1} \approx 10^{40}$ and $N_{2} \approx 10^{40}$, where

\begin{equation}
\label{eq15}
N_{1} = \frac{{e^{2}}}{{Gm^{2}}};\quad N_{2} = \frac{{\left( {{{c} 
\mathord{\left/ {\vphantom {{c} {H}}} \right. \kern-\nulldelimiterspace} 
{H}}} \right)}}{{\left( {{{e^{2}} \mathord{\left/ {\vphantom {{e^{2}} 
{mc^{2}}}} \right. \kern-\nulldelimiterspace} {mc^{2}}}} \right)}}.
\end{equation}

The first number is the ratio of Coulomb force acting between two electrons 
to Newtonian force of their attraction. The second number is the radius of 
the Universe expressed in the units of the classic electron radius. Using 
expressions (\ref{eq15}) for $N_{1} $ and $N_{2} $ and equating $N_{1} $ and $N_{2} 
$, we arrive immediately at approximate relation (\ref{eq14}). 

Thus, the Dirac large number hypothesis can be viewed as a consequence of 
the CGD equations that owes its existence to the conformal symmetry of the 
Universe. 

\subsection*{Example 2}

More realistic is the class of axially symmetric (AXS) solutions of the CGD 
equations. Although no general solutions to the AXS problem for the CGD 
equations have been found, it can be stated almost definitely that in the 
simplest case the AXS solution is determined by two constants: gravitational 
radius $\left( {{{Gm} \mathord{\left/ {\vphantom {{Gm} {c^{2}}}} \right. 
\kern-\nulldelimiterspace} {c^{2}}}} \right)$ and Compton length $\left( 
{{{\hbar}  \mathord{\left/ {\vphantom {{\hbar}  {mc}}} \right. 
\kern-\nulldelimiterspace} {mc}}} \right)$. Reasoning similar to that for 
the spherically symmetric case shows that discontinuity surface radius $a$ 
is

\begin{equation}
\label{eq16}
a = Const \cdot \sqrt {\left( {{{Gm} \mathord{\left/ {\vphantom {{Gm} 
{c^{2}}}} \right. \kern-\nulldelimiterspace} {c^{2}}}} \right) \cdot \left( 
{{{\hbar}  \mathord{\left/ {\vphantom {{\hbar}  {mc}}} \right. 
\kern-\nulldelimiterspace} {mc}}} \right)} = Const \cdot \sqrt 
{\frac{{G\hbar} }{{c^{3}}}} .
\end{equation}

Radius $a$ agrees with the Plank length with an accuracy to a constant. It 
is noticeable that the $a$ is independent of particle mass. In other words, 
the radius of the surface of sewing of two branches of the ASS solution is 
the same in all particles with spin $\hbar /2$. 

Thus, a consequence of the conformal geometrodynamics is the relation among 
gravitational radius, Compton length and Plank length which is well known in 
physics.

\section*{4. A nonperturbative method for estimation of relations between physical 
quantities}

\bigskip

Let us next assume that, besides geometric quantities, a physical situation 
is described by some physical field $\varphi $, which depends on the metric 
tensor and Weyl field in some, maybe complex manner. Let field $\varphi $ be 
correspondent with some physical quantity $\Phi \left( {\varphi}  \right)$.

Denote the physical quantities corresponding to solutions $Large$ and 
$meso$by $\Phi _{Large} $ and $\Phi _{meso} $, respectively. As in the CGD 
equations there is no dimensional constant, all the physical quantities can 
be expressed in terms of length, that is the dimension of Ô is a certain 
degree of length. Therefore preserved quantity $\Phi _{little} $ should be 
related with $\Phi _{Large} $, $\Phi _{meso} $ by the same relation, which 
holds for lengths (radii):

\begin{equation}
\label{eq17}
\left( {S_{little} \cdot \Phi _{little}}  \right) \cdot \left( {S_{Large} 
\cdot \Phi _{Large}}  \right)\sim\left( {S_{meso} \cdot \Phi _{meso}}  
\right)^{2}.
\end{equation}

Quantities $S_{little} $, $S_{meso} $, $S_{Large} $ have a meaning close to 
that of statistical weights (combinatorial multipliers), i.e. methods for 
realization of states $little$, $meso$, $Large$, respectively, while their 
reciprocals have the meaning of probabilities of the relevant states. 

The form of the relation between physical quantities is easy to use, where 
all weights (or probabilities) are combined into a single reduced 
``normalization'' factor. Then the relation can be given by

\begin{equation}
\label{eq18}
\Phi _{little} = \xi \frac{{\Phi _{meso}^{2}} }{{\Phi _{Large}} }.
\end{equation}

Formula (\ref{eq18}), which we will term as Pushkin's relation, provides a basis for 
what follows. In specific cases the exact determination of $\xi _{little} 
,\;\xi _{Large} ,\;\xi _{meso} $ can be complex enough. As for the methods 
for determination of multipliers $\xi _{little} ,\;\xi _{Large} ,\;\xi 
_{meso} $, in his book [9] Pushkin writes that they appeared for three 
principal reasons:

`` a) unit vector enumeration combinatoric analysis; 

\noindent
b) presence of various types of reflective symmetries, including mere 
inversion of one or more spatial unit vectors;

\noindent
c) possibility of exchange of a temporal unit vector for a spatial in one 
signature sector (for example, $t \to ir$and simultaneously $r \to it$).

Inclusion of these symmetries is a simplest method for bundle averaging of 
trajectories (solutions), where the comparison of two solution sets proceeds 
over ``rough'' invariants, which do not distinguish them by these signs. Of 
course, the primary source of these symmetries are topological and 
differential-topological properties of the manifolds under consideration. 
Arithmetically, the summation or averaging over bundle trajectories 
manifests itself, primarily, in simple combinatoric factors like 
$\frac{{1}}{{4!}} = \frac{{1}}{{24}};\;\,\frac{{2}}{{3}}$; etc. in algebraic 
formulas relating solution invariants.''

To be more specific regarding the above quotation note that in some cases 
weight factor $\xi $ in (\ref{eq18}) can take on both very large and vary small 
values. This can happen, when the group of symmetries of states $little$, 
$meso$, $Large$ has a large order. Such was the case, for example, in the 
consideration in [14] of the vacuum solutions corresponding to the lower 
orbit of lattice $E_{8} $; in that case the weight factor was close to the 
order of Weyl group of algebra $E_{8} $, i.e. to $\sim 0.7 \cdot 10^{9}$. 

Clear that all of the aforesaid holds not only in regard to the scalar 
physical field $\varphi $ with a certain Weyl weight $k$, but also the 
physical field of any nature, for example, bispinor field or gauge field 
(like Yang-Mills field). The only constraint is that the physical fields 
should be geometric objects of the Weyl space, in particular, have a certain 
Weyl weight. 

Thus understood physical fields should not, generally speaking, be 
identified with metric $g_{\alpha \beta}  $ and Weyl vector $A_{\alpha}  $. 
The physical fields (bispinor, gauge, etc. fields) are prescribed in the 
Weyl space, in which there are tensor $g_{\alpha \beta}  $ and vector 
$A_{\alpha}  $, whose explicit form is found as a result of solving the CGD 
equations. The physical fields obey their dynamics which can relate in a 
very complex fashion with dynamics of $g_{\alpha \beta}  $and $A_{\alpha}  $ 
fields. Practically, the phenomenological techniques of field description 
that have been elaborated by theoretical physics can be made use of to 
determine it. It should be kept in mind that \underline {whatever solution 
in terms of physical fields we consider, a certain solution}\\
\underline {to the CGD equations exists simultaneously with it 
as well in each}\\
\underline{space-time domain}. 

The parallel existence of a solution in terms of physical fields and a 
solution to the CGD equations in each spatial domain can cause appearance of 
relations between physical quantities, if the assumptions specified in 
subsection 2.2 are fulfilled. 

From the above procedure of Pushkin's formula derivation it follows that 
Pushkin's method under discussion is applicable only to states related by 
inverse conformal transformations. The search for the states of this kind is 
an independent problem. 

We have conducted a search (far from complete) for inverse conformal states. 
The search led us to conclude that Pushkin's method under discussion can be 
attempted to apply to the following processes:

(\ref{eq1}) Decaying of the first excited level of hydrogen atom, i.e. the process 
of transition of the hydrogen atom from state $2p$ to state $1s$.

(\ref{eq2}) Emitting of 21-cm radio line by hydrogen atom.

(\ref{eq3}) Decaying of neutron in free state.

All the three processes belong to a single class characterized with the 
following.

A) The states under consideration are causally connected. That is, they 
refer to processes occurring with one and the same physical object, with at 
least one of the states being quasi-stationary (quasi-stable). In his book 
[9] Pushkin correlates these bound or quasi-bound states with solutions in 
the Euclidean sector of signature $\left( { + + + +}  \right)$. The process 
of the transition from the metastable state to the stable is described by a 
Lorentz metric solution. In the geometrodynamics, the initial and final 
states are related by general (not spherically symmetric) solution in 
signature sector $\left( { - + + +}  \right)$. The solution in signature 
sector $\left( { - + + +}  \right)$ is an ``instanton'' with respect to the 
one in Euclidean sector $\left( { + + + +}  \right)$.

B) The processes under consideration are of dissipative nature, i.e. the 
nature, in which the processes under consideration cannot be explained in a 
natural fashion within the framework of the standard versions of effective 
theories. Note that the CGD equations include the dissipative processes 
without violation of the causality principle. Responsible for the 
dissipative process resulting in the transition from one state to another 
are the conformally invariant interactions. 

Â) The states under consideration are salient physically. The salience 
(fundamentality) of the states means that the states refer to the primary 
elementary particles and nuclei and their lowest energy levels. 

In what follows we consider each of the above processes and show how 
relations between characteristics of the processes can be determined.

\section*{5. Nontrivial examples of Pushkin's method application}

\bigskip

\subsection*{5.1. Width of the first excited level of hydrogen atom}

\bigskip

Let us consider the hydrogen atom and ask ourselves if relation (\ref{eq18}) can be 
used as a basis to obtain an estimator of radiation lifetime $\tau $ of the 
first excited level 2p of the hydrogen atom. In the quantum theory, the 
ground state of atom, its excited state 2p and the process of the transition 
from the excited state to the ground one are considered to obtain the 
estimator. In the energy representation, the excited state is characterized 
with energy $E_{1} $ counted from the ground state and equal energy of 
photons emitted in transition $2p \to 1s$. The lifetime $\tau $is 
correspondent with level width $\Delta E_{1} = \frac{{\hbar} }{{\tau _{1} 
}}$. The ground state is characterized with energy of the atom as a whole, 
in the atom rest frame the energy is $E_{0} = M_{H} c^{2}$. 

As we said earlier, Pushkin correlates the excited and ground states with 
solutions in Euclidean sector of signature $\left( { + + + +}  \right)$. The 
process of the transition from the metastable state to the stable is 
described by the solution with Lorentz metric. In the geometrodynamics, the 
initial and final states are related by general (not spherically symmetric) 
solution in signature sector $\left( { - + + +}  \right)$. For the hydrogen 
atom this is the geometrodynamic representation of photon emission. The 
solution in signature sector $\left( { - + + +}  \right)$ is an 
``instanton'' with respect to the one in Euclidean sector $\left( { + + + + 
} \right)$.

Each of the above characteristics is correspondent with triply connected CGD 
solutions according to (4.1). The characteristics of this connection are 
determined by inequality

\begin{equation}
\label{eq19}
E_{0} > > E_{1} > > \Delta E_{1} .
\end{equation}

In this case formula (\ref{eq18}) will be written as

\begin{equation}
\label{eq20}
\Delta E_{1} = \frac{{E_{1}^{2}} }{{E_{0}} }\xi _{s} ,
\end{equation}

\noindent
and the lifetime of level $2p$ is given by

\begin{equation}
\label{eq21}
\tau = \frac{{\hbar E_{0}} }{{\xi _{s} E_{1}^{2}} }.
\end{equation}

In the substitution of the known values of $\hbar $, $E_{0} $, $E_{1} $ into 
the right-hand side we obtain$\tau = \frac{{5.93}}{{\xi _{s}} } \cdot 10^{ - 
9}s$, while the experimental value is $\tau _{e}^{} = 1.6 \cdot 10^{ - 
9}s$[15]. For $\xi _{s} = 4$ the theoretical value is $\tau = 1.48 \cdot 
10^{ - 9}s$. In our opinion based on quantum mechanics intuition, this value 
of $\xi _{s} $ is correspondent with $S_{0} = 1,S_{1} = 1,S_{\Delta}  = 2$, 
however the current determination of $\xi _{s} $ requires plunging into the 
deep methods of analysis of the geometrodynamics equations which were being 
developed by Pushkin.

Among the three states appearing in the consideration performed for the 
hydrogen atom, one refers to the atom as a whole. The question is 
appropriate: Why on earth does the proton-related term $M_{H} c^{2}$ appear 
in the electron level related estimators? When answering a question of this 
kind, one should keep in mind that the typically used factorization of the 
wave function in the form of product of the wave function of nucleus by the 
wave function of electrons is nothing more than a supposition. In the 
framework of CGD the atom is a connected system, for which there is no wave 
function factorization in the entire space-time domain; the wave function 
can be represented approximately in one spatial domain and cannot in 
another. Therefore, when considering properties of electrons entering into 
the composition of the atom, the use of $M_{H} c^{2}$ can prove quite 
appropriate.

Additional ``food for thought'' is provided by the analysis of the triplet: 
hydrogen, deuterium and tritium. Formula (\ref{eq28}) includes atomic mass. On this 
evidence it may seem that for deuterium the lifetime of the first excited 
state is 2 times as long as that in hydrogen, and in tritium this is 3 times 
as long. It is well known, however, that the lifetime of the first excited 
level of hydrogen, deuterium and tritium atoms is the same. The seeming 
contradiction is removed by the fact that for deuterium the additional 
multiplier 2!=2 due to the proton and neutron transposition should be 
included in the left-hand side. For tritium the multiplier is equal (with 
taking into account the identity of two neutrons) to 3=3!/2!.

\subsection*{5.2. Estimation of hydrogen radio line emission time}

\bigskip

Imagine that the transition of hydrogen atom from state $ \uparrow \uparrow 
$ to state $ \uparrow \downarrow $ proceeds under the action of microwave 
background radiation, which is a heat reservoir (``bath'') for the electron.

We will estimate background radiation energy density $\varepsilon _{\gamma}  
$ assuming that:

1) $10^{9}$ background radiation photons are contained in 1 m$^{3}$.

2) Order-of-magnitude energy of each of the photons is $\frac{{1}}{{2}}kT$.

Assuming $T = 2.73$ K, we obtain that energy density $\varepsilon _{\gamma}  
$ is

\begin{equation}
\label{eq22}
\varepsilon _{\gamma}  = 1.88 \cdot 10^{ - 13}erg/cm^{3}.
\end{equation}

We will determine the width of level $ \uparrow \uparrow $, which will be 
denoted by $\Delta E_{ \uparrow \uparrow}  $, with Pushkin's formula (\ref{eq18}); 
in this case it will take the form

\begin{equation}
\label{eq23}
\Delta E_{ \uparrow \uparrow}  = \left( {{{\left( {\Delta E_{atom}}  
\right)^{2}} \mathord{\left/ {\vphantom {{\left( {\Delta E_{atom}}  
\right)^{2}} {E_{electron}} }} \right. \kern-\nulldelimiterspace} 
{E_{electron}} }} \right).
\end{equation}

Here:

 $\Delta E_{atom} $ is the difference of background radiation energies contained 
in the initial and final states,

 $E_{electron} $ is the background radiation energy contained in the electron 
volume.

 $E_{electron} $ can be estimated as follows:

\begin{equation}
\label{eq24}
E_{electron} = \left( {{{\hbar}  \mathord{\left/ {\vphantom {{\hbar}  {mc}}} 
\right. \kern-\nulldelimiterspace} {mc}}} \right)^{3} \cdot \varepsilon 
_{\gamma}  .
\end{equation}

As for electron Compton length is $\left( {{{\hbar}  \mathord{\left/ 
{\vphantom {{\hbar}  {mc}}} \right. \kern-\nulldelimiterspace} {mc}}} 
\right) = 3.8 \cdot 10^{ - 11}cm$, then for $E_{electron} $ we obtain:

\begin{equation}
\label{eq25}
E_{electron} = 1.03 \cdot 10^{ - 44}\;erg.
\end{equation}

 $\Delta E_{atom} $ will be estimated as

\begin{equation}
\label{eq26}
\Delta E_{atom} = {E}'_{atom} - {E}''_{atom} ,
\end{equation}

\noindent
where ${E}'_{atom} $, ${E}''_{atom} $ are the background radiation energies 
contained in the initial and final states. ${E}'_{atom} $, ${E}''_{atom} $ 
are:

\begin{equation}
\label{eq27}
{E}'_{atom} = \frac{{4}}{{3}}\pi r_{0}^{3} \cdot \varepsilon _{\gamma}  
;\quad {E}''_{atom} = \frac{{4}}{{3}}\pi \left( {r_{0} - \Delta r_{0}}  
\right)^{3} \cdot \varepsilon _{\gamma}  
\end{equation}

Here $r_{0} = \left( {{{\hbar ^{2}} \mathord{\left/ {\vphantom {{\hbar ^{2}} 
{me^{2}}}} \right. \kern-\nulldelimiterspace} {me^{2}}}} \right) = 5.29 
\cdot 10^{ - 9}\;cm$ is the radius of the first Bohr orbit. As for $\Delta 
r_{0} $, it has the meaning of decrease in the radius of the first Borh 
orbit in the transition corresponding to photon emission of wavelength 
$\lambda = 21.1\;cm$ (i.e. hydrogen radio line). The emitted photon energy 
$\hbar \omega $ is determined by formula

\[
\hbar \omega = \left( {{{2\pi \hbar c} \mathord{\left/ {\vphantom {{2\pi 
\hbar c} {\lambda} }} \right. \kern-\nulldelimiterspace} {\lambda} }} 
\right) = 0.929 \cdot 10^{ - 17}Erg.
\]

Then our reasoning is as follows: if electron binding energy $E_{1} = 
13.6\;eV = 2.18 \cdot 10^{ - 11}\;erg$ on the first Bohr orbit is 
correspondent with radius $r_{0} $, then binding energy $E_{1} + \hbar 
\omega $ should be correspondent with radius $r_{0} - \Delta r_{0} $. Whence

\[
r_{0} - \Delta r_{0} = r_{0} \frac{{E_{1}} }{{E_{1} + \hbar \omega} }.
\]

Since $\hbar \omega < < E_{1} $,

\begin{equation}
\label{eq28}
\Delta r_{0} = r_{0} \frac{{\hbar \omega} }{{E_{1}} } = 2.25 \cdot 10^{ - 
15}\;cm.
\end{equation}

For $\Delta E_{atom} $ we find:

\begin{equation}
\label{eq29}
\Delta E_{atom} = 4\pi r_{0}^{2} \left( {\Delta r_{0}}  \right) \cdot 
\varepsilon _{\gamma}  = 1.49 \cdot 10^{ - 43}\;erg.
\end{equation}

The substitution of the determined values of $\Delta E_{atom} $ and 
$E_{electron} $ into the formula for $\Delta E_{ \uparrow \uparrow}  $ 
yields the following:

\[
\Delta E_{ \uparrow \uparrow}  = \left( {{{\left( {\Delta E_{atom}}  
\right)^{2}} \mathord{\left/ {\vphantom {{\left( {\Delta E_{atom}}  
\right)^{2}} {E_{electron}} }} \right. \kern-\nulldelimiterspace} 
{E_{electron}} }} \right) = 2.16 \cdot 10^{ - 42}\;erg
\]

Knowing line width $\Delta E_{ \uparrow \uparrow}  $, determine average 
lifetime $\tau $.

\begin{equation}
\label{eq30}
\tau = \left( {{{\hbar}  \mathord{\left/ {\vphantom {{\hbar}  {\Delta E_{ 
\uparrow \uparrow} } }} \right. \kern-\nulldelimiterspace} {\Delta E_{ 
\uparrow \uparrow} } }} \right) = 4.8 \cdot 10^{14}\;sec = 1.5 \cdot 
10^{7}\;years.
\end{equation}

The evaluated average lifetime $\tau $ is close to the $\tau \approx 
10^{7}\;years$determined from astrophysical considerations (see [16]). This 
result points to both the validity of the physical notions of the background 
radiation role and the validity of Pushkin's formula for this case.

\subsection*{5.3. Estimation of neutron lifetime}

\bigskip

The process of free neutron decay by scheme

\[
n \to p + e + \tilde {\nu} 
\]

\noindent
is governed by electroweak interactions. The reaction goes through 
generation of intermediate charged boson, photons cannot be directly 
involved in this decay. 

However, in the primeval form (before spontaneous break of symmetry) 
electroweak interactions can be formulated in the conformally invariant 
form. Therefore it is of interest to treat this process by Pushkin's method.

Two quantities with energy dimension are associated with the decay process. 
First, quantity

\[
\left( {\Delta M} \right)c^{2} = \left( {M_{n} - M_{p}}  \right)c^{2} = 
1.29\;MeV = 2.06 \cdot 10^{ - 6}\;Erg.
\]

Second, background radiation energy $E_{b - g} $contained in neutron volume

\[
E_{b - g} = \left( {\frac{{\hbar} }{{M_{n} c}}} \right)^{3} \cdot 
\varepsilon _{\gamma}  = \left( {2.08 \cdot 10^{ - 14}\;cm} 
\right)^{3}\left( {1.88 \cdot 10^{ - 13}\;\frac{{Erg}}{{cm^{3}}}} \right) =
\]
\[ 
=1.69 \cdot 10^{ - 54}\;Erg.
\]

Determine level width $\Delta E$ by Pushkin's formula, which in this case is 
given by:

\[
\Delta E = \sqrt {\left( {\Delta M} \right)c^{2} \cdot E_{b - g}}  = \sqrt 
{\left( {2.06 \cdot 10^{ - 6}\;Erg} \right) \cdot \left( {1.69 \cdot 10^{ - 
54}\;Erg} \right)} =
\]
\[
= 1.87 \cdot 10^{ - 30}\;Erg.
\]

This level width is correspondent with level lifetime $T$ of

\[
T = \frac{{\hbar} }{{\Delta E}} = \frac{{\left( {1.04 \cdot 10^{ - 27}\;Erg 
\cdot sec} \right)}}{{\left( {1.87 \cdot 10^{ - 30}\;Erg} \right)}} = 
560\;sec.
\]

The experimental neutron lifetime is $\left( {888 \pm 10} \right)\;sec$. The 
comparison of $T$ with the experimental value shows that Pushkin's method 
does work in this case as well, but its accuracy, as expected, is not high. 
This may be because in the neutron decay the background radiation photons 
play an auxiliary role - as a mechanism of neutron perturbation in free 
state resulting in the neutron decay.

\textbf{5.4 Biological addendum}

\bigskip

Pushkin [9] presents an illustration of a chain of processes, for which the 
conformal geometrodynamics provides tools for construction of solutions. The 
chain begins with elementary particles and ends with global cosmological 
structures, that is begins with the smallest scales and ends with the 
largest. The conformal geometrodynamics symmetry allows the solution 
invariants not only of the nearest, but also of not nearest neighbors in the 
chain to be connected. Therefore, even having no explicit solutions, we can 
ask ourselves about connection of quantities that are far from each other in 
scales. Actually, we have already considered problems of this kind in 
Section 3. 

In this statement a natural question arises: What will the average scale be, 
if for the large scale we take the size of the Universe

 $L_{H} = cH^{ - 1} = 4300Mpc = 1.33 \cdot 10^{28}$ cm,

\noindent
and for the small scale the Plank length 

 $l_{Pl} = \left( {\frac{{\hbar G}}{{c^{3}}}} \right)^{\frac{{1}}{{2}}} = 1.61 
\cdot 10^{ - 33}$ cm.

Then by formula (\ref{eq22}) with $\xi $=1 we obtain the following for the medium 
scale:

 $l_{c} = \sqrt {L_{H} l_{Pl}}  = 4.6 \cdot 10^{ - 3}$ cm. (31)

Surprising as it may seem, the human cell sizes lie precisely in the 5-100µm 
range. Moreover, the spermatozoid and ovicell nucleus sizes range within 
50-60 µm. It turns out that the objects of paramount importance to the 
existence of the mankind as well as all living things are equidistant from 
the ``dangerous'' largest and smallest sizes. 

The journey to biology can be continued, but we will restrict our 
consideration only to one more example. The man consists mainly of water 
molecules. Take for the small size the water molecule diameter

 $l_{H_{2} O} = 3 \cdot 10^{ - 8}$ cm,

\noindent
for the large size the man's medium height 

 $L_{hb} = 160$ cm ,

\noindent
then 

 $l_{c}^{1} = \sqrt {l_{H_{2} O} L_{hb}}  = $ 22 µm.

The small cells are more in number in a body than the large. This estimator 
objectively characterizes the harmonicity (optimality) of the human body 
construction. In this consideration the elephant is definitely overheavy, 
while the mouse is too light.

The examples given in this section are in essence well known. Their 
interpretation becomes more visual, if we write the golden section formula 
in the logarithmical form. Then, for example, relation (31) will become

\[
lnl_{c} = \frac{{1}}{{2}}\left( {lnl_{Pl} + lnL_{H}}  \right),
\]

\noindent
that is on the scale axis the cells of living organisms are located exactly 
in the middle between the Universe and Plank length. Many details of the 
analysis of life and human in the scale of the Universe in the logarithmical 
form are described in ref. [22]. Of course, the subject of the scale harmony 
is quite ancient, a great many studies are devoted to it, and many books are 
written about it. All art and architecture are inconceivable without the 
harmony. There are different approaches to its description. The approach 
discussed by us seminally links the scale harmony of the World with Weyl 
geometry and its conformal symmetry.

\section*{6. Conclusion}

\bigskip

In this paper we made an attempt to represent separate aspects of 
Pushkin's approach to the estimation of fundamental physical constants 
and relations between physical quantities. Pushkin's approach is based on 
the analysis (group, geometrical, functional) of conformal geometrodynamics. 
We restricted our consideration to those estimators which follow immediately 
from CGD equation solution symmetry about the conformal transformations and 
differentiable changes of coordinates. Finest features of the method 
relating to the analysis of symmetry of the conformal quantum field theory 
with the vacuum polarization tensor coinciding with the energy-momentum 
tensor of the CGD equations are therewith left aside. According to Pushkin, 
a symmetry group termed the ``Monster'' group acts in this quantum field 
theory. The Monster group has sufficient ``building material'' 
($\sim 8.08\cdot 10^{53}$ elements) for the calculations with an accuracy competitive 
with that of most precision experiments of the modern physics. Yet, we 
restricted ourselves to the order-of-magnitude estimations. 

This paper is based on several nontrivial rigorously proven facts. The 
proven facts include the presence of a specific symmetry in the stationary 
solutions of the CGD equations which is due to the conformal inverse 
transformations. The symmetry leads to appearance of a connection between 
lengths characterizing the position of the singularities of the internal and 
external solution parts. The connection, in its turn, leads to appearance of 
the relation between physical quantities characterizing either of the above 
solution parts. Besides, in CGD there is a possibility of self-consistent 
description of decay processes, that is the possibility to describe the 
transition between two (quasi-) bound states using the general solution to 
the CGD equations. This solution also leads to appearance of the relation 
between the physical characteristics of the process and states which has the 
form of the generalized golden ratio. 

This paper uses the above facts and some assumptions to consider two 
commonly known ("large number hypothesis'' and relation among gravitational 
radius, Plank radius, and Compton length) and three nontrivial examples of 
using Pushkin's relation (\ref{eq18}) between physical quantities. These examples, 
of course, do not exhaust all possible ``golden section'' relations. In this 
paper it was important to us to demonstrate that certain physical CGD models 
stand behind Pushkin's relation (\ref{eq18}), that the relations are not merely a 
result of fitting or guessing. The examples given in the paper, as we 
believe, have accomplished this task. 

Speaking about the significance of the relations between physical quantities 
that follow from the conformally inverse symmetry, two aspects may be 
mentioned.

- In some cases these relations can allow us to obtain a quantitative 
estimation of a physical quantity, like this was done for the lifetime of 
the first excited state of hydrogen atom: if we had not known that time, we 
could have estimated it. 

- Dimensionless relations between physical quantities can serve prompts to 
reveal the implication of interrelations between effects and processes. 
Attention is drawn to this aspect in many papers (see, e.g., [17]-[21]). For 
example, the standard model of electroweak interactions cannot be viewed as 
a fundamental theory of the interactions, as the model includes 
phenomenological parameters that are not deduced from it itself. The 
dimensionless relations between physical quantities, as it seems to us, can 
be demanded both in improvement of this model and in development of a 
higher-level theory.

\bigskip

\section*{References}

\bigskip

[1] A.V.Pushkin. \textit{'Monstrous Moonshine' and Physics.} Proceedings of 
the Second International A.D. Sakharov Conference on Physics. World 
Scientific, 316-319 (1996). 

[2] M.V.Gorbatenko, A.V.Pushkin. \textit{On the correspondence between 
tensors and bispinors. Part VI. (The nature of background radiation)}. // 
VANT-TPF. \textbf{3. P.} 3-9 (2004). 

[3] M.V.Gorbatenko, A.V.Pushkin. \textit{Dynamics of the linear affine 
connectedness space and conformally invariant extension of Einstein 
equations.} // VANT-TPF. \textbf{2/2.} P. 40-46 (1984).

[4] M.V.Gorbatenko, A.V.Pushkin. \textit{Thermodynamic analysis of Weyl 
geometry base geometrodynamics equations. //} VANT-TPF. Issue 2, pp. 17 - 
21, 1992.

[5] M.V.Gorbatenko, A.V.Pushkin, H.-J.Schmidt. \textit{On a relation between 
the Bach equation and the equation of geometrodynamics.} General Relativity 
and Gravitation. \textbf{34}, No. 1, 9-22 (2002). 

[6] M.V.Gorbatenko, A.V.Pushkin. \textit{Conformally Invariant 
Generalization of Einstein Equation and the Causality Principle.} General 
Relativity and Gravitation. \textbf{34}, No. 2, 175-188 (2002).

[7] M.V.Gorbatenko, A.V.Pushkin. \textit{Addendum. Conformally Invariant 
Generalization of Einstein Equation and the Causality Principle.} General 
Relativity and Gravitation. \textbf{34}, No. 7, 1131-1133 (2002).

[8] A.V.Pushkin. \textit{On construction of a system of axially symmetric 
stationary solutions to geometrodynamics equations. Part I.} VANT-TPF. Issue 
\textbf{3,} pp. 53-69 (2001).

[9] A.V.Pushkin. \textit{Geometrodynamics. A program of development of 
algorithms for construction of analytical solutions to equations describing 
two-dimensional and three-dimensional continuum motions.} Sarov. RFNC-VNIIEF 
Publishing Polygraphic Complex (2005).

[10] V.Canuto,P.J.Adams, S.-H.Hsieh, and E.Tsiang. \textit{Scale-covariant 
theory of gravitation and astrophysical applications}. // Physical Review D, 
Vol. 16, Number 6. P.p. 1643-1663 (1977).

[11] E.Fairchaild. \textit{Gauge theory of gravitation.} // Phys. Rev. 
\textbf{D}14, 384 (12976).

[12] Brans C., Dicke R.H. // Phys. Rev. \textbf{124}. P.925 (1961).

[13] M.V.Gorbatenko. // Advances in Mathematics Research. Vol. 6. 
Editors:\textbf{} Oyibo, Gabriel. Nova Science Publishers, Inc. (2005).

[14] M.V.Gorbatenko, A.V.Pushkin. // GRG. \textbf{37}, No. 10, 1705-1718 
(2005).

[15] F.Bethe, E.Solpiter. \textit{Quantum mechanics of atoms with one and 
two electrons.} Moscow, GIFML Publishers (1960).

[16] R.L.Sorochenko. \textit{21-cm radio line of hydrogen}// Space Physics. 
Small Encyclopedia. Moscow. Sovyetskaya Entsiklopediya Publishers (1986).

[17] M.A.Markov. \textit{Selected works}. V.1. Moscow. Nauka Publishers 
(2000). V.2. Moscow. Nauka Publishers (2001).

[18] L.B.Okun. Fundamental Constants of Nature. // arXiv:hep-ph/9612249.

[19] V.E.Shemi-zadeh. \textit{Coincidence of large numbers, exact value of 
cosmological parameters and their analytical representation}. // 
arXiv:gr-qr/0206084.

[20] D.A.Varshalovich, A.V.Ivanchik, A.Yu.Potekhin. \textit{Fundamental 
physical constants: Are their values identical in different space-time 
domains?} // Zhurnal Tekhnicheskoy Fiziki. V.69, Issue 9. Pp.1-5 (1999).

[21] Saibal Ray, Utpal Mukhopadhyay, Partha Pratim Ghosh. \textit{Large 
Number Hypothesis}. // arXiv:0705.1836 [gr-qc].

[22] S.I.Sukhonos, N.P.Tretyakov. \textit{The man in the scale of the 
Universe}. Moscow. Novy Tsentr Publishers (2004)

\end{document}